\documentclass[ 
showpacs,
showkeys,
amsmath,amssymb,
preprint
]{revtex4-1}

\usepackage{diagbox}
\usepackage{graphicx}
\usepackage{dcolumn}
\usepackage{bm}
\usepackage[usenames,dvipsnames]{color}
\usepackage{ulem}

\usepackage{amsmath,esint}

\usepackage{enumitem}
\usepackage{listings}

\definecolor{mygreen}{RGB}{28,172,0} 
\definecolor{mylilas}{RGB}{170,55,241}
\definecolor{MyDarkGreen}{rgb}{0.0,0.4,0.0}

\begin{document}

\global\long\def\R{\mathbb{R}}

\global\long\def\Rn{\mathbb{R}^{n}}
\global\long\def\Zn{\mathbb{Z}^{n}}

\global\long\def\eps{\varepsilon}

\global\long\def\ue{u^{\eps}}

\global\long\def\bQ{\boldsymbol{Q}}
\global\long\def\cS{\mathcal{S}}

\global\long\def\cL{\mathcal{F}}

\global\long\def\cF{\mathcal{L}}
\global\long\def\d{\mathrm{d}}

\title{Markov models from the Square Root Approximation of the Fokker-Planck equation: calculating the grid-dependent flux}

\author{Luca Donati}
 \affiliation{Department of Biology, Chemistry, Pharmacy, Freie Universit\"at Berlin, Takustra\ss e 3, D-14195 Berlin, Germany}
\author{Marcus Weber}%
\affiliation{Zuse Institute Berlin, Takustr. 7, 14195 Berlin, Germany}
\author{Bettina G. Keller}
 \email{bettina.keller@fu-berlin.de} 
 \affiliation{Department of Biology, Chemistry, Pharmacy, Freie Universit\"at Berlin, Takustra\ss e 3, D-14195 Berlin, Germany}

\date{\today}

\begin{abstract}
Molecular dynamics are extremely complex, yet understanding the slow components of their dynamics is essential to understanding their macroscopic properties.
To achieve this, 
one models the molecular dynamics as a stochastic process and analyses the dominant eigenfunctions of the associated Fokker-Planck operator, or of closely related transfer operators. 
So far, the calculation of the discretized operators requires extensive molecular dynamics simulations. 
The Square-root approximation of the Fokker-Planck equation is a method to calculate transition rates as a ratio of the Boltzmann densities of neighboring grid cells times a flux, and can in principle be calculated without a simulation.
In a previous work we still used molecular dynamics simulations to determine the flux.
Here, we propose several methods to calculate the exact or approximate flux for various grid types,
and thus estimate the rate matrix without a simulation.
Using model potentials we test computational efficiency of the methods, and the accuracy with which they reproduce the dominant eigenfunctions and eigenvalues. 
For these model potentials, rate matrices with up to $\mathcal{O}(10^6)$ states can be obtained within seconds on a single high-performance compute server if regular grids are used.
\end{abstract}

\maketitle
\section{Introduction}
The dynamics of molecular systems is astonishingly complex. 
Only a small fraction of their high-dimensional state space is actually accessible at room temperature. 
Yet finding out which regions of the state space are accessible, requires sophisticated computer simulations, i.e. molecular dynamics (MD) simulations.
Molecular dynamics can be very sensitive to small changes in some variables of the  system or the environment, but can also be remarkably robust with respect to changes in other variables.
Humanly understandable models of the molecular dynamics are therefore essential for the elucidation of complex molecular systems.
Markov state models (MSMs) represent the conformational dynamics of molecular system as transition probabilities between states in the conformational space
\cite{Schuette1999b, Swope2004,Buchete2008,Keller2010, Prinz2011, Wang2018b}.
From the dominant eigenvectors and eigenvalues of the transition matrix $\mathbf{T}(\tau)$, 
one can deduce a wealth of useful information on the molecular system, 
such as the long-lived conformations, the dynamic processes that govern the dynamic equilibrium between them, transition networks and pathways in these networks, and one can quantify the sensitivity of experimental observables with respect to the dynamic processes \cite{Prinz:2011b, Husic:2018}.
MSMs are now a well-established and valuable tool for the elucidation of large molecular systems, and in particular biomolecular systems \cite{Voelz2010b, Stanley2014, Bowman2015, Plattner2015, Zhang2016, Witek2016, keller2018}. 
In the construction of MSMs, one assumes that the molecular dynamics is a stochastic process. 
The time-evolution of the probability density is governed by the associated
Fokker-Planck equation, or equivalently: the infinitesimal generator of the stochastic process $\mathcal{Q}$.
By formally integrating the Fokker-Planck equation one obtains a transfer operator, whose discretized version is the MSM transition matrix $\mathbf{T}(\tau)$.
The matrix elements of $\mathbf{T}(\tau)$ can conveniently be estimated from MD simulations as correlation functions.
On the other hand, this means that the accuracy of the MSM stands and falls with the quality of this simulation.
Because MD simulations are costly and slow to converge,
enhanced sampling techniques have been developed to speed up the exploration of state space and the convergence of ensemble averages \cite{Pietrucci:2017,Zuckerman:2017,Valsson:2016,Abrams:2014}. 
With recently developed dynamic reweighting methods one can additionally recover the correlation functions and thus the MSM of the unbiased system from these biased simulations \cite{Chodera:2011, Rosta2014, Donati2018, Kieninger2020}.
But despite enhanced sampling techniques, there is usually no way to be certain whether an MD simulation has explored all of the accessible state space, and even assessing whether the sampling within the explored state space has converged can be difficult \cite{Grossfield:2018,Zuckerman:2011}.
Thus, there is ample motivation to investigate avenues to obtain a MSM of a molecular system without generating a MD simulation.
Square Root Approximation (SqRA) is a technique that approximates the Fokker-Planck equation by a rate matrix \cite{Lie2013, Donati2018b}.
Given a discretization of the state space, the rate from cell $\Omega_i$ to cell $\Omega_j$ is 
\begin{eqnarray*}
Q_{ij,\,\mathrm{adjacent}} 
&=& \Phi \,\frac{\mathcal{S}_{ij}}{\mathcal{V}_i}\sqrt{\frac{\pi(x_j)}{\pi(x_i)}} \, ,
\end{eqnarray*}
where $\Phi \,\frac{\mathcal{S}_{ij}}{\mathcal{V}_i}$ is the flux of the probability density through the intersecting surface $\mathcal{S}_{ij}$ in the absence of any potential energy function, 
$\mathcal{V}_i$ is the volume of cell $\Omega_i$, and $\pi(x_i)$ and $\pi(x_j)$ are the Boltzmann densities at the centers of cell $\Omega_i$ and $\Omega_j$, respectively. 
We recently derived the SqRA for $N_D$-dimensional systems by exploiting Gauss's flux theorem, and showed that for infinitely small grid cells the geometric average of the Boltzmann weights converges to the Smoluchowski diffusion equation, i.e.~the Fokker-Planck equation associated to overdamped Langevin dynamics \cite{Heida2018, Donati2018b}.
Previously an analogous formula for one-dimensional systems has been derived from the one-dimensional Smoluchowski equation \cite{Bicout1998} and using the maximum caliber (maximum path entropy) approach \cite{Dixit2015, Stock2008, Otten2010}.
In addition the geometric average of the Boltzmann weights has been used as reweighting factor in the dynamic histogram analysis method (DHAM) to reweight transition probabilities \cite{Rosta2014}.
The SqRA opens up a way to calculate the transition rates without having to resort to rare-event simulations, at least for systems with not too many degrees of freedom.
The ratio of the Boltzmann densities $\pi(x_i)/\pi(x_j)$ can be readily calculated from the potential energy function. The grid volume $\mathcal{V}_i$ and the intersecting surface $\mathcal{S}_{ij}$ can be calculated from the discretization of the state space. 
However, how to best calculate $\Phi \,\frac{\mathcal{S}_{ij}}{\mathcal{V}_i}$ is an open question.
In our previous work \cite{Donati2018b}, we assumed that the factor is $\frac{\mathcal{S}_{ij}}{\mathcal{V}_i}$ is constant for all grid cells. This is true for hyper-cubic grids and approximately true for Voronoi grids with very small grid cells. 
We then estimated the factor $\Phi \frac{\mathcal{S}_{ij}}{\mathcal{V}_i}$ by comparing the rate matrix to a MSM transition matrix, 
the construction of which required an MD simulation.

In this contribution, we derive the exact expression for $\Phi$ from the equation of the overdamped Langevin dynamics with constant potential, and show that it depends on the diffusion constant and on the discrete Laplace operator.
We then compare several methods to calculate $\frac{\mathcal{S}_{ij}}{\mathcal{V}_i}$ for different types of discretizations.
For regular grids, this ratio can be calculated analytically. 
For Voronoi grids, we use the quickhull algorithm \cite{Barber1996} to calculate $\frac{\mathcal{S}_{ij}}{\mathcal{V}_i}$ numerically,
and we approximate the ratio by interpolating between all neighbors of  the cell $\Omega_i$ \cite{Oostendorp1989}.
We additionally propose a method to calculate $\Phi \frac{\mathcal{S}_{ij}}{\mathcal{V}_i}$ by comparing to the analytically known transition probability of a Wiener process (i.e~diffusion at a constant potential energy function).
With theses methods, we can construct the rate matrix without any MD simulation.
We test the methods on model potentials with respect to computational efficiency, the dimensionality of the systems, and accuracy of the resulting rate matrix.


\section{Theory}
We consider a system of $n_p$ particles that move in the three-dimensional Cartesian space, 
i.e.~in a state space with $N_D = 3 n_p$ dimensions: $\Omega \subset \mathbb{R}^{N_D}$.
Its dynamics is described by the overdamped Langevin dynamics:
\begin{equation}
    \mathrm{d}x(t) = -\,  \xi^{-1} \mathbf{M}^{-1} \, \nabla V(x(t)) \mathrm{d}t + \sigma \mathrm{d} B(t) \, ,
   \label{eq:sde}
\end{equation}
where
$x(t) \in\Omega$ is the state vector at time $t$, 
$\xi$ is a friction parameter with units of 1/s, 
$\mathbf{M}$ is a diagonal $3 n_p \times 3 n_p$-mass matrix,
$\mathbf{M}^{-1}$ is its inverse,
$V(x)$ is the potential energy function,
and $B(t)$ is an $N_D$-dimensional Wiener process
scaled by the diagonal matrix $\sigma = \sqrt{2 k_B T \xi^{-1}  \mathbf{M}^{-1}}$,
%
%
where 
$T$ is the temperature, and $k_B$ is the Boltzmann constant.
Eq.~\ref{eq:sde} generates a Markovian, ergodic and reversible process \cite{Schuette1999b, Risken1989}. 
The time-evolution of the associated probability density $\rho(x,t)$
is given by the following Fokker-Planck equation
\begin{eqnarray}
\partial_t \rho(x,t)  
&=& \frac{\sigma^2}{2}  \Delta \rho(x,t) + \nabla \left(\rho(x,t) \cdot \xi^{-1}  \mathbf{M}^{-1} \nabla V(x) \right)\cr
&=& \mathcal{Q}\rho(x,t)\,, 
\label{eq:FP}
\end{eqnarray}
which is also known as the Smoluchowski diffusion equation.
The symbol $\nabla$ denotes the gradient of a function $f: \mathbb{R}^n \rightarrow \mathbb{R}$, and $\Delta = \nabla \cdot \nabla$ is the corresponding Laplacian.
The factor in front of the Laplacian can be interpreted as the matrix of the diffusion coefficients $\mathbf{D} = \frac12 \sigma^2$, which are assumed to be independent of the particle positions \cite{Risken1989}.
Eq.~\ref{eq:FP} introduces the Fokker-Planck operator $\mathcal{Q}$.
$\mathcal{Q}$ can also be interpreted as the infinitesimal generator of a transfer operator (or propagator) with lag time $\tau$: $\mathcal{T}(\tau) = \exp \left(\mathcal{Q}\tau \right)$. 
The operator $\mathcal{T}(\tau)$ propagates $\rho(x,t)$ forward in time by a time interval $\tau$: $\mathcal{T}(\tau)\rho(x,t) = \rho(x,t+\tau)$.
The stationary solution of eq.~\ref{eq:FP} is the Boltzmann density
\begin{eqnarray}
    \pi(x) &=& 
    \frac{\exp \left(-\frac{1}{k_BT} V(x) \right)}{Z}
\label{eq:stationary_density}    
\end{eqnarray}
where $Z=\int_{\Omega} \exp \left(-\frac{1}{k_BT} V(x) \right) \,\mathrm{d}x$ is the classical
partition function, 
i.e. $\partial_t \pi(x) = \mathcal{Q}\pi(x) = 0$.
%

%
\subsection{Square Root Approximation}
\label{sec:sqra}
The square root approximation (SqRA) of the infinitesimal generator is a method to
discretize $\mathcal{Q}$, and to calculate the corresponding matrix elements \cite{Lie2013, Donati2018b}.
We will briefly review its derivation in the following section.
Consider a disjoint decomposition of the state space $\Omega$ into $N$ Voronoi cells $\Omega_i$, 
such that $\Omega = \cup_{i=1}^N \Omega_i$.
The characteristic function associated to each Voronoi cell $\Omega_i$ is
\begin{eqnarray}
    \chi_i(x) &=&
    \begin{cases}
        1   &\mathrm{if} \, x \in \Omega_i \cr
        0   &\mathrm{otherwise} \, ,
    \end{cases}
\label{eq:characteristic_fct}    
\end{eqnarray}
We introduce the following scalar product $\langle u,v\rangle_{\pi} = \int_{\Omega} u(x) \, v(x)\, \pi(\mathrm{d}x) = \int_{\Omega} u(x) \, v(x)\, \pi(x) \mathrm{d}x$. For disjoint sets, the Galerkin discretization of $\mathcal{Q}$ is computed via $Q=(\left<\chi_j|\chi_i \right>_\pi)_{ij}^{-1}(\left<\chi_j|\mathcal{Q}\chi_i\right>_\pi)_{ij}$ which reduces to
\begin{eqnarray}
    Q_{ij} &=&
    \frac{1}{\pi_i} \left< \chi_j | \mathcal{Q} \chi_i \right>_{\pi} \, ,
\label{eq:Q_matrix_element}      
\end{eqnarray}
if we use eq.~\ref{eq:characteristic_fct} as ansatz functions.
The term $\pi_i =  \left< \chi_i | \chi_i \right>_{\pi}= \int_{\Omega_i} \pi(x)\mathrm{d}x$ denotes the stationary probability of cell $\Omega_i$.

Eq.~\ref{eq:Q_matrix_element} defines a $N \times N$
transition rate matrix $\mathbf{Q}$ with elements $Q_{ij}$, 
where $Q_{ij}$, for $i\not= j$, denotes the rate from cell $\Omega_i$ to cell $\Omega_j$.
The discretization is analogous to the discretization of the transfer operator in 
the derivation of Markov State Models (MSMs) \cite{Schuette1999b, Prinz2011}, which yields a transition matrix $\mathbf{T}(\tau)$.
Just as in MSMs, the state space $\Omega$ is usually so high-dimensional that solving the integral in eq.~\ref{eq:Q_matrix_element} is not a viable option.
However, in contrast to MSMs, the numerator in eq.~\ref{eq:Q_matrix_element} cannot be estimated from correlation functions obtained by simulating the stochastic process in  eq.~\ref{eq:sde} \cite{Prinz2011, Nuske2014}. 
The square root approximation provides a solution to this impasse, 
which neither requires solving the high-dimensional integral nor sampling the
stochastic process. 
The derivation starts by noting that for time-homogeneous processes
the rate matrix and the transition matrix are related by
$\mathbf{Q}:=\left.\frac{\partial \mathbf{T}(\tau)}{\partial \tau}\right\vert_{\tau=0}$. 
For infinitesimally small lag times $\tau$, the transition rates between cells which do not share a common boundary is certainly zero. 
Thus, we can set the rate matrix elements for non-adjacent cells to
\begin{eqnarray}
    Q_{ij} &=& 0 \qquad \mbox{if $i\ne j$, and $\Omega_i$ is not adjacent to $\Omega_j$} \, .
\end{eqnarray}
Because the matrix elements $T_{ij}(\tau)$ represent transition probabilities, we can use the Gauss theorem to show that the rate matrix elements for adjacent cells satisfy \cite{Lie2013, Donati2018b}
\begin{equation}
Q_{ij} = \frac{1}{\pi_i} \oint_{\partial\Omega_i \partial\Omega_j} \Phi(z) \,  \pi(z) \mathrm{d} S(z) \, , 
\label{gauss_1}
\end{equation}
where $\oint$ denotes a surface integral. Furthermore, $\partial \Omega_i \partial \Omega_j$ is the common surface between the cell $\Omega_i$ and $\Omega_j$.  
$\Phi(z)=\delta_{\Omega_i=\Omega_j}\mathbf{v}\cdot \mathbf{n}$ denotes the flux of the configurations $z$ through the  surface $\partial\Omega_i \partial\Omega_j$. 
The vector $\mathbf{v}$ is the velocity field associated to the time-dependent probability density.
This is analogous to the fluid velocity in fluid dynamics, which describes the velocity of a small element of fluid such that the mass is conserved.

To approximate the surface integral in eq.~\ref{gauss_1}, we introduce the first of two assumptions of SqRA:
\begin{enumerate}
\item The flux does not depend on the position in state space: 
$\Phi(x) = \Phi$.
Then 
\begin{eqnarray}
Q_{ij} &=& \frac{1}{\pi_i} \Phi \oint_{\partial\Omega_i \partial\Omega_j}  \,  \pi(z) \mathrm{d} S(z) \, .
\label{eq:SqRA_rate01}        
\end{eqnarray}
\end{enumerate}
The remaining surface integral in eq.~\ref{eq:SqRA_rate01} 
represents the stationary density on the intersecting surface $\partial \Omega_i \partial \Omega_j$.  
To approximate it, we formulate our second assumption:
\begin{enumerate}
\setcounter{enumi}{1}
\item Each cell is small such that the potential energy $V(x)$ is almost constant within the cell: $V(x) |_{\Omega_i} \approx V_i$.
\end{enumerate}
It follows that the stationary density $\pi(x)$, and, by extension, also the time-dependent density $\rho(x,t)$, 
is constant within a given cell $\Omega_i$.
The continuous and the discretized probabilities are related by
\begin{eqnarray}
    \pi_i &=& \int_{\Omega_i} \pi(x)\, \mathrm{d}x \,\approx\, \pi(x_i) \mathcal{V}_i \cr
    \rho_i(t) &=& \int_{\Omega_i} \rho(x,t)\, \mathrm{d}x \,\approx\, \rho(x_i,t) \mathcal{V}_i 
\label{eq:discretizedDensity}    
\end{eqnarray}
where $\int_{\Omega_i}1\, \mathrm{d}x=\mathcal{V}_i$ is the volume of the cell $\Omega_i$, 
and in particular we have
\begin{eqnarray}
    \pi(x_i) &=& \frac{1}{Z}  \exp\left(-\frac{1}{k_BT} V_i \right) \, ,
\end{eqnarray}
where $x_i$ is the center of $\Omega_i$.
Likewise, we can assume that the potential energy
function on $\partial \Omega_i \partial \Omega_j$ is essentially constant, and that it can be approximated by some average of $V_i$ and $V_j$.
We choose the arithmetic mean $ V(x) |_{\partial\Omega_i \partial \Omega_j} \approx \frac{V_i + V_j}{2}$, 
because for this type of mean-value calculation one can show that the resulting discretized operator $\mathbf{Q}$ converges to the Fokker-Planck-operator $\mathcal{Q}$ in the limit of infinitesimally small cells \cite{Heida2018, Donati2018b}.
The surface integral in eq.~\ref{eq:SqRA_rate01} then becomes 
\begin{eqnarray}
    \oint_{\partial\Omega_i \partial\Omega_j}  \, \pi(z) \mathrm{d} S(z)
    =  \oint_{\partial\Omega_i \partial\Omega_j}  \, \frac{1}{Z}\exp \left(-\frac{1}{k_BT} \frac{V_i + V_j}{2} \right) \mathrm{d} S(z)
    &=& \mathcal{S}_{ij} \sqrt{\pi(x_i) \pi(x_j)}\, ,
    \label{eq:surface_integral}
\end{eqnarray}
where $\oint_{\partial\Omega_i \partial\Omega_j}\,1\,\mathrm{d} S(z) =\mathcal{S}_{ij}$ is the area of the intersecting surface.
Note that an arithmetic mean of the potential energy function results in a geometric mean of the stationary densities: $\sqrt{\pi(x_i) \pi(x_j)}$.
With this appoximation of the surface integral and with eq.~\ref{eq:discretizedDensity}, we obtain the following expression for rates between adjacent cells (eq.~\ref{gauss_1})
\begin{eqnarray}
Q_{ij,\,\mathrm{adjacent}} 
&=& \frac{1}{\pi_i} \,  \Phi \,\mathcal{S}_{ij} \sqrt{\pi(x_i) \pi(x_j)}
=    \Phi \,\frac{\mathcal{S}_{ij}}{\mathcal{V}_i}\sqrt{\frac{\pi(x_j)}{\pi(x_i)}}  \, , 
\label{eq:rate_adjacent}
\end{eqnarray}
and the following rate matrix 
\begin{eqnarray}
Q_{ij} &=& 
\begin{cases}
\Phi \,\frac{\mathcal{S}_{ij}}{\mathcal{V}_i}\sqrt{\frac{\pi(x_j)}{\pi(x_i)}}   
                                            &\mbox{if  $i\ne j$, and  $\Omega_i$ is adjacent to $\Omega_j$}  \\
0                                           &\mbox{if  $i\ne j$, and  $\Omega_i$ is not adjacent to $\Omega_j$} \\
-\sum_{j=1, j\ne i}^n Q_{ij}                            &\mbox{if } i=j \, .
\end{cases} 
\label{eq:rate_matrix_01}
\end{eqnarray}
This is the SqRA of the Fokker-Planck operator $\mathcal{Q}$.
Note that in our previous publication \cite{Donati2018b}, we did not write the factor $\mathcal{S}_{ij}/\mathcal{V}_i$ explicitly, 
because we assumed that it is approximately the same for all pairs of adjacent cells and can be incorporated into $\hat{\Phi} = \Phi \,\frac{\mathcal{S}_{ij}}{\mathcal{V}_i}$.

The discretization of the Fokker-Planck equation (eq.~\ref{eq:FP}) then is
\begin{eqnarray}
    \partial_t \boldsymbol{\rho}^{\top}(t) &=& \boldsymbol{\rho}^{\top}(t) \mathbf{Q}
\label{eq:FP_discretized01}    
\end{eqnarray}
where $\boldsymbol{\rho}(t)$ is the vector-representation of the continuous probability density $\rho(x,t)$ with elements 
$\rho_i(t) 
= \int_{\Omega_i} \rho(x,t) \, \mathrm{d}x 
= \int_{\Omega} \rho(x,t) \chi_i(x) \, \mathrm{d}x$, 
and $\boldsymbol{\rho}^{\top}(t)$ denotes the transpose of $\boldsymbol{\rho}(t)$.
Eq.~\ref{eq:FP_discretized01} can be rewritten as an evolution equation for the individual vector elements 
\begin{eqnarray}
    \partial_t \rho_j(t) 
    &=& \sum_{i=1}^N \rho_i(t) Q_{ij}
    = \left[\sum_{i=1, i \ne j }^N \rho_i(t) Q_{ij}\right] +\rho_j(t) \left[-\sum_{k=1, k \ne j }^n Q_{jk}\right]
\end{eqnarray}
which is often written more concisely as a master equation
\begin{eqnarray}
    \partial_t \rho_j(t)
    &=& \sum_{i \sim j}  \left[\rho_i(t) Q_{ij} - \rho_j(t)  Q_{ji}\right] \, ,
\label{eq:FP_discretized02}       
\end{eqnarray}
where  $\sum_{i\sim j}$ denotes the sum over all adjacent cells $\Omega_i$ of cell $\Omega_j$.

The great appeal of the SqRA of the Fokker-Planck operator is that, apart from the grid-dependent flux $\Phi \,\frac{\mathcal{S}_{ij}}{\mathcal{V}_i}$, 
it only requires the Boltzmann-density $\pi(x_i)$ at the cell centers (eq.~\ref{eq:rate_matrix_01}), 
which are readily available from the potential energy surface of the system.
In principle, no time-series are required to calculate the rate matrix. 
The challenge lies in estimating $\Phi \,\frac{\mathcal{S}_{ij}}{\mathcal{V}_i}$.
In the following, we introduce 
two different approaches to calculate $\Phi \,\frac{\mathcal{S}_{ij}}{\mathcal{V}_i}$ that do not rely on a realization of eq.~\ref{eq:sde}.

%
%
\subsection{$\Phi \,\frac{\mathcal{S}_{ij}}{\mathcal{V}_i}$ by discretizing the Laplacian}
\label{sec:Theory_Laplacian}
If the flux $\Phi$ does not depend on the potential energy function (assumption 1), 
one should be able to determine $\Phi$ by analyzing the overdamped Langevin dynamics on a constant potential $V(x) = \mathrm{const.}$, 
\begin{equation}
    \mathrm{d}x_t = \sigma \mathrm{d} B_t \, ,
\label{eq:sde_constV}    
\end{equation}
and the associated Fokker-Planck equation
\begin{eqnarray}
\partial_t  \rho(x,t) &=&  \frac{\sigma^2}{2}\Delta  \rho(x,t) = \mathcal{Q} \rho(x,t) \, .
\label{eq:FP_constV}    
\end{eqnarray}
This has two advantages. First, the differential operator in eq.~\ref{eq:FP_constV} essentially consists of the Laplacian, whose discretization is known. 
Second, the stationary density (eq.~\ref{eq:stationary_density}) of this process is constant, which simplifies the expression for the rates (eq.~\ref{eq:rate_adjacent}).

Applying the Gauss theorem, the Laplacian of the probability density $\rho (x,t)$ over a small region with volume $\mathcal{V}$ and surface $\mathcal{S}$, is written as \cite{Arfken2001} 
\begin{equation}
    \Delta \rho (x,t) = \lim_{\mathcal{V} \rightarrow 0} \frac{1}{\mathcal{V}} \oint_{\mathcal{S}} \nabla \rho(z,t) \cdot \mathbf{n} \, \mathrm{d}S(z) \, ,
\end{equation}
where $\mathbf{n}$ is the unit vector orthogonal to the surface $\mathcal{S}$.
It follows, that on a Voronoi tessellation of the space, the discrete Laplacian on a small Voronoi cell $\Omega_i$ is \cite{Sukumar2003, Kil2011}
\begin{equation}
    \left.\Delta \rho(x,t) \right\vert_{x = x_{j}} = \frac{1}{\mathcal{V}_j} \sum_{i\sim j} \left.\nabla \rho(x,t)\right\vert_{x = x_j} \cdot \mathbf{n}_{ji} \mathcal{S}_{ji} \, .
    \label{eq:discLap1}
\end{equation}
The term $\left.\nabla \rho(x,t)\right\vert_{x = x_j} \cdot \mathbf{n}_{ji}$ is the gradient in the direction $j\rightarrow i$ (directional derivative), 
which can be approximated by the finite difference
\begin{equation}
    \left.\nabla \rho(x,t) \right\vert_{x = x_j} \cdot \mathbf{n}_{ji}   \approx \frac{\rho(x_i, t) - \rho(x_j, t)}{h_{ji}} \, ,
\label{eq:rho_finite_difference}    
\end{equation}
where 
$h_{ji} = x_j - x_i$
is the distance between the centers of the cells $\Omega_j$ and $\Omega_i$.
Inserting this finite difference into eq.~\ref{eq:discLap1} yields
\begin{equation}
     \left.\Delta \rho(x,t) \right\vert_{x = x_j}= \frac{1}{\mathcal{V}_j} \sum_{i\sim j}  \frac{\rho(x_i, t) - \rho(x_j, t)}{h_{ji}} \mathcal{S}_{ji} \, .
    \label{eq:discLap2}
\end{equation}
Assuming that the density $\rho(x,t)$ is approximately constant within cell $\Omega_i$ (assumption 2),  
we have $\rho_i(t) \approx \int_{\Omega_i}\rho(x_i, t) \,\mathrm{d}x = \rho(x_i,t) \mathcal{V}_i$. 
Substituting $\rho(x_i,t) = \frac{\rho_i(t)}{\mathcal{V}_i}$ in eq.~\ref{eq:discLap2} and inserting into eq.~\ref{eq:FP_constV} yields
\begin{eqnarray}
\partial_t  \frac{\rho_j(t)}{\mathcal{V}_j} 
&=&  \frac{\sigma^2}{2} \frac{1}{\mathcal{V}_j} \sum_{i\sim j}  \frac{\frac{\rho_i(t)}{\mathcal{V}_i} - \frac{\rho_j(t)}{\mathcal{V}_j}}{h_{ij}} \mathcal{S}_{ij} \, ,
\end{eqnarray}
and we obtain the the discrete Fokker-Planck equation (eq.~\ref{eq:FP_constV}) at constant potential
\begin{eqnarray}
\partial_t  \rho_j(t)
&=&  \frac{\sigma^2}{2}  \sum_{i\sim j}  \frac{\frac{\rho_i(t)}{\mathcal{V}_i} - \frac{\rho_j(t)}{\mathcal{V}_j}}{h_{ij}} \mathcal{S}_{ij} 
= \sum_{i\sim j} \frac{\sigma^2}{2}  \frac{1}{h_{ij}}\frac{\mathcal{S}_{ij}}{\mathcal{V}_i}\rho_i(t) - \frac{\sigma^2}{2}  \frac{1}{h_{ij}}\frac{\mathcal{S}_{ij}}{\mathcal{V}_j}\rho_j(t)
\label{eq:discFPE}
\end{eqnarray}
Comparing eq.~\ref{eq:discFPE} to the master equation (eq.~\ref{eq:FP_discretized02}) and to the definition of rates between adjacent cell within the SqRA (eq.~\ref{eq:rate_adjacent})
we obtain the following equality
\begin{eqnarray}
Q_{ij,\,\mathrm{adjacent}}  =  \Phi \,\frac{\mathcal{S}_{ij}}{\mathcal{V}_i} &=& \frac{\sigma^2}{2}  \frac{1}{h_{ij}}\frac{\mathcal{S}_{ij}}{\mathcal{V}_i}
\label{eq:gridFlux}
\end{eqnarray}
where we used that $\sqrt{\frac{\pi(x_j)}{\pi(x_i)}} =1 $ at constant potential. 
Thus,
\begin{equation}
    \Phi = \frac{\sigma^2}{2 h_{ij}} \,.
\label{eq:flux}    
\end{equation}{}
We have obtained an analytical expression for  $\Phi$ between adjacent cells that only depends on the distance $h_{ij}$ between the cell centers. %
Appendix B contains an alternative derivation of  eq.~\ref{eq:flux} using Fick's first law of diffusion.

The rate matrix (eq.~\ref{eq:rate_matrix_01}) can now be written more concretely as 
\begin{eqnarray}
Q_{ij} &=& 
\begin{cases}
\frac{\sigma^2}{2} \,\frac{1}{h_{ij}}\,\frac{\mathcal{S}_{ij}}{\mathcal{V}_i}\sqrt{\frac{\pi(x_j)}{\pi(x_i)}}   
                                            &\mbox{if  $i\ne j$, and  $\Omega_i$ is adjacent to $\Omega_j$}  \\
0                                           &\mbox{if  $i\ne j$, and  $\Omega_i$ is not adjacent to $\Omega_j$} \\
-\sum_{j=1, j\ne i}^n Q_{ij}                            &\mbox{if } i=j \, .
\end{cases} 
\label{eq:rate_matrix_02}
\end{eqnarray}
Section \ref{sec:methods} introduces formulas to evaluate $\frac{1}{h_{ij}}\,\frac{\mathcal{S}_{ij}}{\mathcal{V}_i}$ for various grid types.

%

%
\subsection{$\Phi \,\frac{\mathcal{S}_{ij}}{\mathcal{V}_i}$ by analyzing the transition probability density}
Our starting point is again eq.~\ref{eq:rate_matrix_01}, and we introduce a third assumption:
\begin{enumerate}
\setcounter{enumi}{2}
    \item The volumes of all cells are approximately equal ($\mathcal{V}_i \approx \mathcal{V}, \forall\, \Omega_i$), 
    and the intersecting surfaces areas are approximately equal for all adjacent cells 
    ($\mathcal{S}_{ij} \approx \mathcal{S}, \forall\, \Omega_i \mbox{ adjacent to } \Omega_j$).
\end{enumerate}
In this case, the factor 
$\Phi \frac{\mathcal{S}_{ij}}{\mathcal{V}_i} \approx \Phi \frac{\mathcal{S}}{\mathcal{V}} = \Phi_{\mathrm{grid}}$ 
has approximately the same value for each pair of adjacent cells. 
$\Phi_{\mathrm{grid}}$ is thus a flux value which is characteristic for a given grid rather than for a specific pairs of cells. 
This is the assumption we used in ref.~\onlinecite{Donati2018b}.

Every grid can be represented as an unweighted graph, in which nodes correspond to the grid cells $\Omega_i$, 
and two nodes are connected by an edge if the corresponding grid cells are adjacent.
At constant potential, the rate matrix (eq.~\ref{eq:rate_matrix_01}) can then be written as the Laplacian matrix of the graph $\mathbf{L}$ multiplied by the grid flux
\begin{eqnarray}
    \mathbf{Q} &=& -\Phi \frac{\mathcal{S}}{\mathcal{V}} \mathbf{L} = -\Phi_{\mathrm{grid}}\mathbf{L}\, ,
\label{eq:rate_matrix_03}    
\end{eqnarray}
where the Laplician matrix of the graph is defined as
\begin{eqnarray}
L_{ij} &=& 
\begin{cases}
-1   &\mbox{if  $i\ne j$, and  $\Omega_i$ is adjacent to $\Omega_j$}  \\
0   &\mbox{if  $i\ne j$, and  $\Omega_i$ is not adjacent to $\Omega_j$} \\
-\sum_{j=1, j\ne i}^n L_{ij}                            &\mbox{if } i=j \, .
\end{cases} 
\label{eq:adjacency_matrix_01}
\end{eqnarray}
Note that $\mathbf{L} = \mathbf{D}-\mathbf{A}$, 
where $\mathbf{A}$ is the adjacency matrix of the graph, 
with elements $A_{ij}=1$ if $\Omega_i$ and $\Omega_j$ are neighbors, and $A_{ij}=0$ otherwise.
$\mathbf{D}$ is the degree matrix of the graph, 
a diagonal matrix whose diagonal entries contain the number of neighbors for each cell, i.e.~$D_{ii} = \sum_{j}A_{ij}$.
The transition matrix $\mathbf{T}(\tau)$ and the rate matrix $\mathbf{Q}$ are related by
\begin{equation}
    \mathbf{T}(\tau) = \exp(\tau\, \mathbf{Q}) =  \exp\left(-\tau\, \Phi_{\mathrm{grid}} \mathbf{L}\right)\,.
\label{eq:transitionMatrix}    
\end{equation}
If one knows the transition probability of a $T_{ij}(\tau)$ of single pair of adjacent cells at constant potential, one can calculate $\Phi_{\mathrm{grid}}$ by comparing $T_{ij}(\tau)$ to the matrix element 
$\left[\exp\left(-\tau\, \Phi_{\mathrm{grid}}\mathbf{L}\right)\right]_{ij}$.
The transition probability is defined as the integral transition probability density $p(x,y,\tau)$ over the initial and final cell

\begin{eqnarray}
    T_{ij}(\tau) &=& \frac{1}{\pi_i} \int_{\Omega_i}\int_{\Omega_j} p(x, y, \tau) \pi(x) \,\mathrm{d}x \,\mathrm{d}y \, ,
\label{eq:transProb01}
\end{eqnarray}
where $\pi(x)$ is the unconditional probability density of finding the system at point $x$ at time $t$, and
$p(x,y,\tau)$ is the conditional probability density of finding the system in $y\mathrm{d}y$ at time $t+\tau$ given that it started in point $x$ at time $t$.
%
%
In ref.~\onlinecite{Donati2018b} we obtained the transition probability by constructing a MSM based on a simulation of the dynamic process.

Here, we propose a different approach. 
We again use the idea that the flux, and by extension $\Phi_{\mathrm{grid}}$, does not depend on the potential energy function. 
Therefore, it can  be determined from an overdamped Langevin dynamics at constant potential energy (eq.~\ref{eq:sde_constV}), for which the transition probability density is:

\begin{eqnarray}
    p(x, y, \tau) &=& \left(\sqrt{\frac{1}{2\pi\sigma^2 \tau}}\right)^{N_D} \exp\left(-\frac{(y-x)^2}{2\sigma^2 \tau} \right) \, ,
\label{eq:transProbDensity}    
\end{eqnarray}
where $N_D$ is the dimension of the state space.
If the cells are small (assumption 2), the distance from any point in cell $\Omega_i$ to any other point in $\Omega_j$ is approximately equal to the distance of the
centers of the two cells, i.e. $y -x \approx x_j - x_i$ for all $x \in \Omega_i, y \in \Omega_j $.
With this assumption eq.~\ref{eq:transProb01} becomes
\begin{eqnarray}
    T_{ij}(\tau) 
    &=& \frac{1}{\pi_i}  \left(\sqrt{\frac{1}{2\pi\sigma^2 \tau}}\right)^{N_D} \exp\left(-\frac{h_{ij}^2}{2\sigma^2 \tau} \right) \int_{\Omega_i}\pi(x) \,\mathrm{d}x \, \int_{\Omega_j} 1 \mathrm{d}y  \cr
    &=& \left(\sqrt{\frac{1}{2\pi\sigma^2 \tau}}\right)^{N_D} \exp\left(-\frac{h_{ij}^2}{2\sigma^2 \tau} \right)\mathcal{V}  \, ,
\label{eq:transProb02}
\end{eqnarray}
where $h_{ij} = x_j - x_i$ is the distance between the centers of $\Omega_i$ and $\Omega_j$.
Besides $N_D$, $\sigma$, and $h_{ij}$, one only needs the average cell volume $\mathcal{V}$ to calculate  $T_{ij}(\tau)$. The lag time $\tau$ can in principle be chosen freely. 

Now that we have a closed-form approximation for $T_{ij}(\tau)$ at constant potential, we can use eq.~\ref{eq:transitionMatrix} to calculate $\Phi_{\rm grid}$.
Because the Laplacian matrix $\mathbf{L}$ is not invertible, we cannot determine $\Phi_{\rm grid}$ by rearranging eq.~\ref{eq:transitionMatrix}. 
Instead we use $\Phi$ as a parameter that minimizes the difference between $T(\tau)$ and $\exp\left(-\tau\, \Phi \mathbf{L}\right)$, i.e.~we minimize the function:
\begin{equation}
f(\Phi) = \left(T_{ij}(\tau) - \left[\exp\left(-\tau\, \Phi \mathbf{L}\right)\right]_{ij}\right)^2 \, .
\label{eq:fluxViaTransProb01}	
\end{equation}
and $\Phi_{\mathrm{grid}} = \underset{\Phi}{\arg\max}  f( \Phi)$.
This approach requires calculating the matrix exponential of the potentially large but sparse matrix $\mathbf{L}$ and is tested in the result section.
It is tempting to avoid the minimization of $f(\Phi)$ (eq.~\ref{eq:fluxViaTransProb01}) by approximating the matrix exponential as a truncated Taylor series, and solving for $\Phi$. 
Mathematically this is possible. 
But the resulting equation for $\Phi_{\rm grid}$ is a poor approximation to the true value of $\Phi_{\rm grid}$, and we do not recommend using this approach. 
Appendix \ref{app:FluxViaTaylor} discusses the details.

\section{Methods}
\label{sec:methods}

In section \ref{sec:Theory_Laplacian} we showed that the grid-dependent flux factor can be expressed in terms of the known parameters $\sigma$ and $h_{ij}$:
$\Phi \frac{\mathcal{S}_{ij}}{\mathcal{V}_i} = \frac{\sigma^2}{2 h_{ij}} \frac{\mathcal{S}_{ij}}{\mathcal{V}_i}$. 
In this section, we summarize methods to evaluate $\mathcal{S}_{ij}/\mathcal{V}_i$ for different grid types.

%
%
\paragraph{Arbitrary grid / exact method.}
"The Quickhull Algorithm" \cite{Barber1996}, implemented in the MATLAB function "convhulln()", computes the convex hull of a set of multidimensional points and can be used to numerically calculate the surface $\mathcal{S}_{ij}$ and the volume $\mathcal{V}_i$ of an arbitrarily shaped cell. 
We will call this the ``exact method'', because it directly calculates $\frac{\mathcal{S}_{ij}}{\mathcal{V}_i}$ 
without any assumptions on the grid geometry. 
However, the algorithm requires not only the centers of the cells, but also the vertices of each cell, 
which makes it computationally expensive for high-dimensional spaces.

%
%
\paragraph{Hyper-rectangular grid.}
On a (hyper-)rectangular grid, the ratio between interface surface and cell volume  is simply given by the cell length in direction $i \rightarrow j$, i.e~$\frac{\mathcal{S}_{ij}}{\mathcal{V}_i} = \frac{1}{h_{ij}}$, and 
\begin{eqnarray}
Q_{ij,\,\mathrm{adjacent}, \mathrm{rectangular}}  &=&  \frac{\sigma^2}{2 h_{ij}^2} \,\sqrt{\frac{\pi(x_j)}{\pi(x_i)}}   \, .
\label{eq:rate_adjacent_rectangular}
\end{eqnarray}
Note that on a (hyper-)cubic grid $h_{ij}=h$ is the same in all grid dimensions, while for a one-dimensional grid one obtains the equation derived in ref.~\onlinecite{Bicout1998} from the one-dimensional reaction-diffusion equation.
Appendix \ref{app:convergenceToFP} shows that eq.~\ref{eq:rate_adjacent_rectangular} converges to the Fokker-Planck equation in the limit of infinitesimally small cells \cite{Donati2018b, Heida2018}. 
%

%
%
\paragraph{Hexagonal grid.}
The apothem $a$ of a cell is the distance from the cell center to one of the midpoint of its sides.
On a two-dimensional hexagonal grid, $a= h/2$, where $h=h_{ij}$ is the distance between cell centers.
Using the apothem we can calculate the intersecting surface, which is equal to the length of each side of the hexagon 
$\mathcal{S}_{ij}= \frac{2}{\sqrt{3}} a = \frac{h}{\sqrt{3}}$, 
Thus, $\frac{\mathcal{S}_{ij}}{\mathcal{V}_i} = \frac{2}{3h_{ij}}$, and the rate between adjacent hexagonal cells is
\begin{eqnarray}
Q_{ij,\,\mathrm{adjacent}, \mathrm{hexagonal}}  &=&  \frac{\sigma^2}{3 h_{ij}^2} \,\sqrt{\frac{\pi(x_j)}{\pi(x_i)}}  \, .
\label{eq:rate_adjacent_hexagonal}
\end{eqnarray}

%
%
\paragraph{Voronoi grid via the neighbors-method.}
On arbitrary Voronoi grids, several methods \cite{Reuter2009} have been proposed to approximate $\frac{\mathcal{S}_{ij}}{\mathcal{V}_i}$.
For example, from the Taylor expansion of a function on an irregular mesh, the rate between adjacent cells can be expressed as \cite{Oostendorp1989} 
\begin{equation}
        Q_{ij,\,\mathrm{adjacent}, \mathrm{Voronoi}} =  \frac{\sigma^2}{2}  \frac{4}{n_i \, \bar{h_{i}} \, h_{ij} } \, \sqrt{\frac{\pi(x_j)}{\pi(x_i)}} \, ,
\label{eq:rate_adjacent_Voronoi02}    
\end{equation} 
where $n_i$ is the number of neighbors of the cell $\Omega_i$, and $\bar{h_{i}}$ is the average distance between the cell $\Omega_i$ and all the neighbors.

%
%
\begin{table}[h]
\begin{tabular}{llll}
\cline{1-4}
    \textbf{Method}                         &\multicolumn{1}{l}{\textbf{Grid}} 
    &\multicolumn{1}{l}{\textbf{$\Phi\frac{\mathcal{S}_{ij}}{\mathcal{V}_i}$}}  & \multicolumn{1}{l}{\textbf{Eq.}} \\ 
\cline{1-4}   
\multicolumn{4}{c}{Exact ratio of intersecting surface area and cell volume} \\
\cline{1-4}
exact        & arbitrary  & $\frac{\sigma^2}{2} \,\frac{1}{h_{ij}}\,\frac{\mathcal{S}_{ij}}{\mathcal{V}_i}$                                                                             & eq.~\ref{eq:rate_matrix_02}                           \\
rectangular  & hyper-cube & $\frac{\sigma^2}{2 h_{ij}^2} $                                                                                                                               & eq.~\ref{eq:rate_adjacent_rectangular}                \\
hexagonal    & 2D-hexagonal  & $\frac{\sigma^2}{3 h_{ij}^2}  $                                                                                                                             & eq.~\ref{eq:rate_adjacent_hexagonal}                  \\
\cline{1-4}   
\multicolumn{4}{c}{Approximate ratio of intersecting surface area and cell volume} \\
\cline{1-4}
neighbors       & Voronoi  & $\frac{\sigma^2}{2}  \frac{4}{n_i \, \bar{h_{i}} \, h_{ij} } $                                                                                              & eq.~\ref{eq:rate_adjacent_Voronoi02}          \\
\cline{1-4}
\multicolumn{4}{c}{Comparison to transition probability} \\
\cline{1-4} \\[-0.5cm]
minimization & arbitrary  & $\underset{\Phi}{\operatorname{argmin}} \left(T_{ij}(\tau) - \left[\exp\left(-\tau\, \Phi \mathbf{L}\right)\right]_{ij}\right)^2$ & eq.~\ref{eq:fluxViaTransProb01}             \\   
\cline{1-4}
\end{tabular}
\caption{Methods to calculate $\Phi\frac{\mathcal{S}_{ij}}{\mathcal{V}_i}$ in the square-root approximation.}
\label{tab:methods}
\end{table}

%
%
\paragraph{Method overview}
Tab.~\ref{tab:methods} summarizes the methods that are now at our disposal to evaluate $\Phi\frac{\mathcal{S}_{ij}}{\mathcal{V}_i}$.
In the following analysis, we will compare the eigenvalues $\kappa_i$ and left eigenvectors $\mathbf{l}_i$ 
\begin{eqnarray}
    \mathbf{l}_i^{\top}\mathbf{Q} &=& \kappa_i \mathbf{l}_i^{\top}
\end{eqnarray}
of rate matrices $\mathbf{Q}$ constructed using eq.~\ref{eq:rate_matrix_03} in combination with the methods in Tab.~\ref{tab:methods}. 
Note that eq.~\ref{eq:rate_matrix_03} implies that the row-sums of each of these matrices $\mathbf{Q}$ are zero, 
consistent with eq.~\ref{eq:rate_matrix_02}.

\section{Results and discussion}
\subsection{Computational efficiency}
%
%

\begin{figure*}[!ht]
  \begin{center}
  \includegraphics[scale=1]{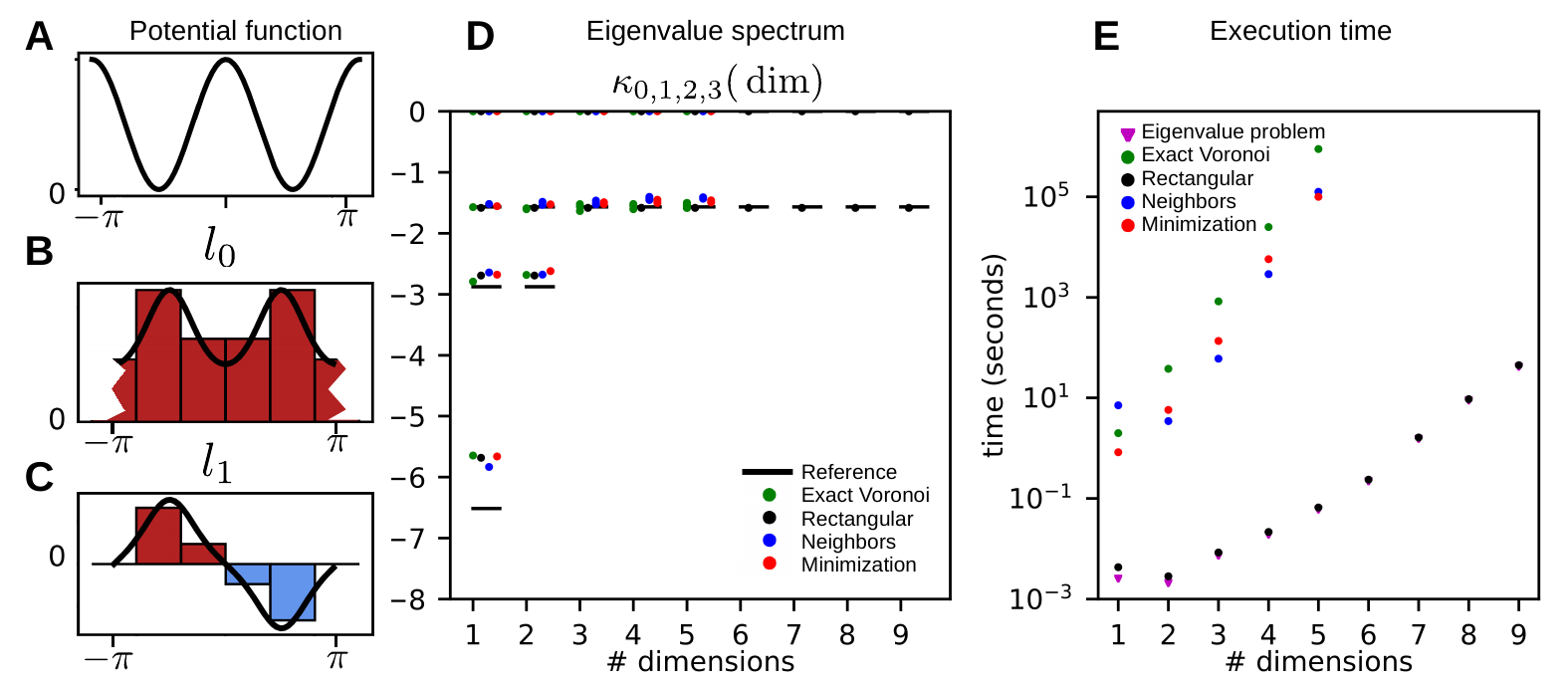}
    \caption{Computational efficiency. (A) One-dimensional periodic potential function, and corresponding first (B) and second (C) eigenfunction of the Fokker-Planck operator $\mathcal{Q}$
    (black lines) compared to the eigenvectors of the corresponding rate matrix $\mathbf{Q}$ (histograms).
    The fringes at $-\pi$ and $\pi$ indicate that this crosses the periodic boundary.
    (D) First four eigenvalues for each $N_D$-dimensional system;
    (E) Execution time for each $N_D$-dimensional system. 
    }
   \label{fig:benchark}
  \end{center}
\end{figure*}

The usefulness of the SqRA critically depends on how many dimensions $N_D$ the dynamical system in eq.~\ref{eq:sde} may have, before the calculation of $\mathbf{Q}$ via eq.~\ref{eq:rate_matrix_01} becomes computationally intractable.
$\mathbf{Q}$ is a $N \times N$ square matrix, where $N$ is the number of cells $\Omega_i$.
If each dimension of the dynamical system is discretized into $N_{\mathrm{bins}}$, the number of cells is given as
$N = N_{\mathrm{bins}}^{N_D}$,
i.e $N$ grows exponentially with the number of dimensions $N_D$.
Thus, even for low-dimensional systems we have to construct a sparse but extremely high-dimensional rate matrix $\mathbf{Q}$. 
To compare the computational efficiency of the methods to estimate $\Phi\frac{\mathcal{S}_{ij}}{\mathcal{V}_i}$,
we devised a model system, which consists of particles of mass $m=1$ $\mathrm{kg}$ moving in  a $N_D$-dimensional Cartesian space  according to eq.~\ref{eq:sde} with $\xi =$ 1 s$^{-1}$ and $\sigma = 2.2$   $\mathrm{J^{\frac{1}{2}} \, kg^{-\frac{1}{2}} \, s^{-\frac{1}{2}}}$.
The potential energy function consists of uncoupled terms for each Cartesian coordinate $x_i$
\begin{equation}
V(x_1,...,x_{N_D}) = \sum_{i=1}^{N_D} V_i(x_i) ,
\end{equation}
defined on the domain
$
\Omega = \lbrace (x_1,...,x_{N_D}) : -\pi < x_i \leq \pi \ \mathrm{for} \ i = 1,...,N_D \rbrace 
$.
%
%
We applied periodic boundary conditions in each direction, 
and the one dimensional potential energy term
\begin{equation}
V_i(x_i) = \frac{1}{2} k_i \left(1 + \cos(m_i \cdot x_i - x_{0i}) \right) \, 
\label{eq:1Dpotential}
\end{equation}
is $2\pi$-periodic in direction $i$.
The parameter $k_i$ is the force constant, $m_i$ is the multiplicity and describes the number of barriers and wells of the function, and $x_{0i}$ is the phase.
For each direction $i$, we used the same triplet of parameters $k_i = 2$ $\rm kg \, s^{-1}$, $m_i=2$ and $x_{0i} = 0$ rad. 
Fig.~\ref{fig:benchark}-A shows the potential for the 1-dimensional system, which is a periodic double well potential.
For an $N_D$-dimensional system, the potential has $2^{N_D}$ wells in the $N_D$-dimensional space.
Eq.~\ref{eq:1Dpotential} mimics the function that governs torsion angles in MD force fields. 
By choosing a Cartesian space with periodic boundary conditions rather then an actual torsion angle, we avoid any complications that arise from the coordinate transformation to the torsion angle space, and a volume element in $\Omega$ is simply given as 
$\mathrm{d}V = \mathrm{d}x_1\mathrm{d}x_2 \dots \mathrm{d}x_{N_D}$.
For the system with $N_D=1$, we constructed a reference solution with 
$N_{\mathrm{bins}} = 60$ bins using the method "rectangular".
The leading eigenvalues of $\mathbf{Q}$ are
$\kappa_0=0$,
$\kappa_1=-1.56$;
$\kappa_2=-2.87$;
$\kappa_3=-6.51$.
Fig.~\ref{fig:benchark}-B shows the eigenvector $l_0$, which corresponds to the stationary distribution. 
Fig.~\ref{fig:benchark}-C  shows the eigenvector $l_1$, which represents a transition 
between the regions $[0, \, \pi)$ and $[\pi,0)$.

We next scanned the number bins between 2 and 60 to find the coarsest possible discretization that still yields accurate results for the dominant processes of the one-dimensional system.
For $N_{\mathrm{bins}}=5$, $\kappa_1 = -1.58$, which is 1.2\% lower than the reference value; while $\kappa_2 = -2.69$ and $\kappa_3=-5.68$ are respectively 6.2 \% and 11.4 \% higher than the reference values. 
A smaller number of bins yield considerable deviations from the reference value.
Fig.~\ref{fig:benchark}-B and \ref{fig:benchark}-C show the approximation of the two 
leading eigenfunctions with $N_{\mathrm{bins}}=5$.
In spite of the very low resolution of the eigenvector, we can identify the two peaks corresponding to the two wells of the potential. 
We constructed grids for up to $N_D=9$ dimensions. 
For the hypercubic grids, we discretized each dimension into $N_{\mathrm{bins}} = 5$ equally-sized bins, 
where the distance between two adjacent states is $h = 2\pi/5 \approx 1.26$.
For the Voronoi grids, we discretized each dimension into $N_{\mathrm{bins}} = 5$ bins of random size.
The number of states $N$ are:
$N_D=1$: 5,
$N_D=2$: 25,
$N_D=3$: 125,
$N_D=4$: 625,
$N_D=5$: 3,125,
$N_D=6$: 15,625,
$N_D=7$: 78,125,
$N_D=8$: 390,625, and
$N_D=9$: 1,935,125 states.
Likewise the memory size of the corresponding rate matrices grows exponentially with $N_D$. For the case with $N_D=9$, the full matrix $\mathrm{Q}$ occupies more than 28 TB of memory, but its corresponding sparse matrix is just 578 MB, which is manageable by modern computers.
We included the following methods in this scan: ``rectangular'' on a hypercubic grid, and ``exact'', ``neighbors'', and ``minimization'' on a Voronoi grid.
The method ``hexagonal" is excluded, 
because a hexagonal grid can only be constructed on a two-dimensional Cartesian space.
The method ``exact'' for the hypercubic grid is not shown explicitly, because it is identical to the method ``rectangular".
For each rate matrix, we calculated the four leading eigenvalues. 
For this system the second eigenvalue has a degeneracy equal to the number of dimensions.
Because they perfectly overlap, there appear to be less eigenvalues for the higher-dimensional systems in fig.~\ref{fig:benchark}-D.
All four methods yield eigenvalues that are in excellent agreement with the reference solution. 
Thus, at least at this level of discretization, approximating the ratio $\frac{\mathcal{S}_{ij}}{\mathcal{V}_i}$ by these methods does not introduce an error of relevant magnitude. 
%

%
However, the computational cost varies drastically between the methods (fig.~\ref{fig:benchark}-E).
Three separate tasks go into calculating the eigenvectors and eigenvalues of $\mathbf{Q}$: 
($i$) constructing the adjacency matrix of the grid from which the Laplacian matrix of the grid $\mathbf{L}$ can then be calculated, 
($ii$) calculating $\Phi\frac{\mathcal{S}_{ij}}{\mathcal{V}_i}$ using one of the four methods, and
($iii$) calculating the dominant eigenvalue-eigenvector pairs for the resulting matrix $\mathbf{Q}$.
The most efficient method is ``rectangular'' on a hypercubic grid, for which we could calculate rate matrices for up to nine dimensions on a server with a Intel Xeon CPU (E5-2690 v3 @ 2.60 GHz) and 160 GB of RAM.
Using MATLAB, the execution time was 45 seconds.
We provide an example script 
On hypercubic grids,  the distance $h_{ij}$ between neighboring cells is a constant, and the factor $\Phi\frac{\mathcal{S}_{ij}}{\mathcal{V}_i} = \frac{1}{2}\frac{\sigma^2}{h^2}$ can be calculated at negligible cost. 
Moreover, one can build adjacency matrices and construct the matrix $\mathbf{Q}$ very efficiently using sparse matrices and the Kronecker product (see supplementary material).
Consequently, approximately the 90 \% of the computational time is used up by the third task: solving the eigenvalue problem (fig.~\ref{fig:benchark}-E, magenta triangles).
The time to solve the eigenvalue problem primarily depends on the dimension of the matrix $\mathbf{Q}$,~i.e. the number of cells $N$.
It does not depend on the method of computing the flux, and it only weakly depends on the type of grid. 
Thus, the computational cost, that is displayed as magenta triangles in fig.~\ref{fig:benchark}-E 
is part of every calculation in fig.~\ref{fig:benchark}-E.
Note that the execution time depends on the algorithm used to solve the eigenvalue problem. The MATLAB function "eigs()" permits to provide the number of eigenvalues-eigenvectors to be calculated. This is particularly useful when one is interested only in the slowest dominant processes, which are associated to the largest eigenvalues.
All three methods to construct the rate matrix on a Voronoi grid are orders of magnitude slower 
than the ``rectangular'' method for hypercubic grids, 
because building adjacency matrices for a Voronoi discretization is computationally difficult and costly.
We constructed the adjacency matrices $\mathbf{A}$ using an algorithm based on linear programming as suggested in ref.~\onlinecite{Lie2013}.
Note that for Voronoi grids, the computational cost of diagonalizing the rate matrix (magenta triangles) makes up only a small fraction of the total calculation. 
Among the methods for a Voronoi grid, the ``exact'' method (green dots in fig.~\ref{fig:benchark}-D) 
is about an order of magnitude more expensive than the methods ``neighbors'' or ``minimization'' (blue and red dots in fig.~\ref{fig:benchark}-D), because it requires the exact calculation of cell volumes. 
Using the same computer as for the ``rectangular'' method, we were able to build the rate matrix of the five dimensional system.
The calculation took $8.9\times 10^5$ s, corresponding to more than ten days of calculations.
However, the ``exact'' method is slightly more accurate then the other three methods for Voronoi grids. 
The methods ``neighbors" and the ``minimization" slightly overestimate the eigenvalues, 
but the execution time reduced to $1.2\times 10^5$ s (33.3 h) and $1.1\times 10^5$ s (30.5 h), respectively.

%
\subsection{Accuracy}
\label{sec:accuracy}
%

\begin{figure*}[!ht]
  \begin{center}
  \includegraphics[scale=1]{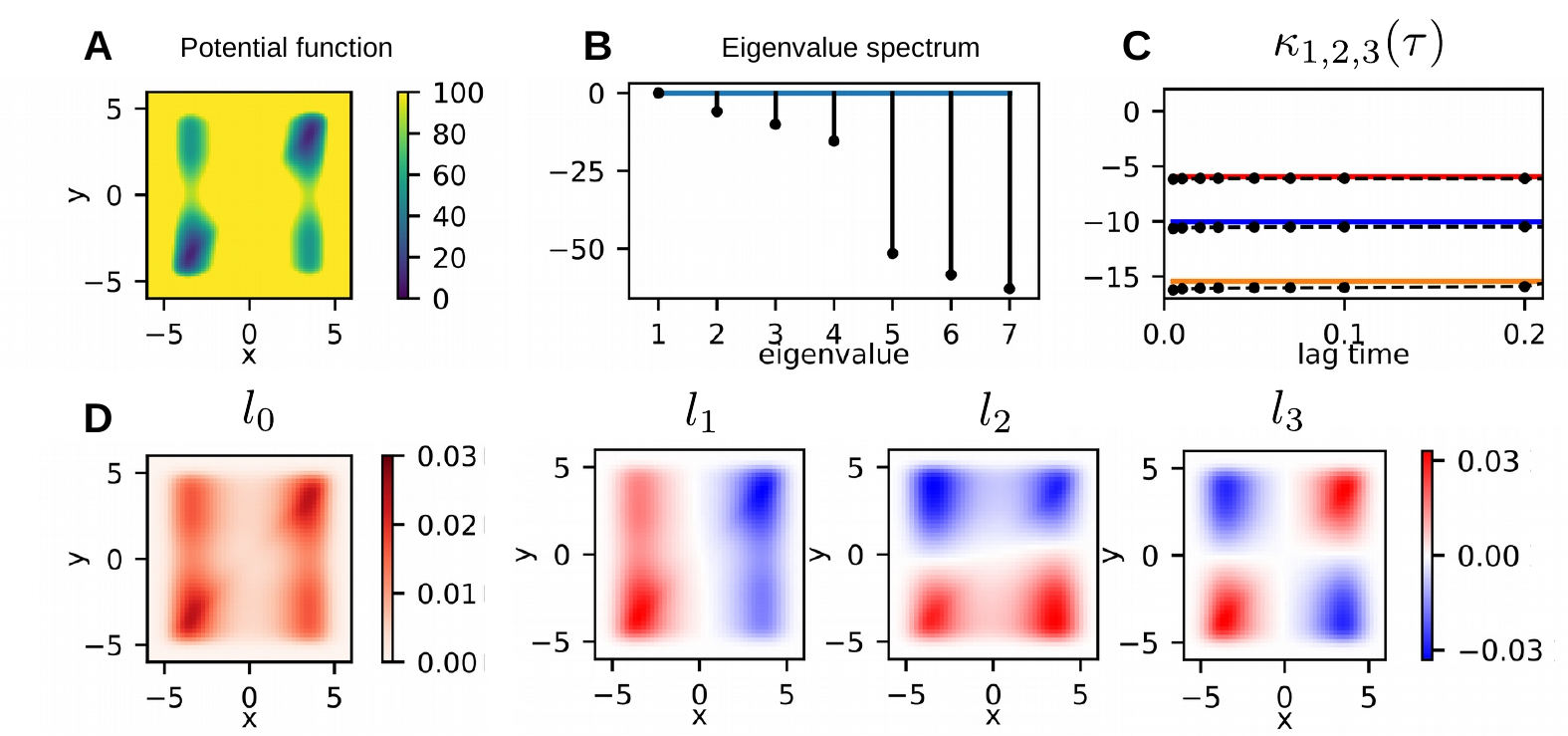}
    \caption{Two dimensional diffusion process in a four well potential. (A) Potential energy function. (B) Eigenvalue spectrum of the corresponding Fokker-Planck operator $\mathcal{Q}$. (C) First three eigenvalues of the infinitesimal generator as function of the lag time $\tau$: SqRA (solid line),  MSM (black dots). (D) First four eigenfunctions of 
    $\mathcal{Q}$.}
   \label{fig:2D_reference}
  \end{center}
\end{figure*}

Next, we test wether the methods in Tab.~\ref{tab:methods} differ in their accuracy. 
We consider a particle of mass $1$ $\mathrm{kg}$ which moves on a two-dimensional Cartesian space 
according to eq.~\ref{eq:sde} with $\xi =$ 1 s$^{-1}$ and 
$\sigma = 15 \,\mathrm{J^{\frac{1}{2}} \, kg^{-\frac{1}{2}} \, s^{-\frac{1}{2}}}$.
The potential energy function is
\begin{equation}
V(x,y) = k_1 (y^2 - a_1^2)^4 + k_2  (x^2 - a_2^2)^2 + \frac{k_{12}}{\sqrt{(x - y)^2 + c^2}} \, ,
\label{eq:pot4well}
\end{equation}
with the parameters $k_1=0.003$, $a_1=3.3$, $k_2=1$, $a_2=3.5$, $k_{12}=-50$ and $c=1$.
This potential is composed of a one-dimensional term which describes a slow double-well dynamics along the $x$ axis, a one-dimensional term which describes a fast double-well dynamics along the $y$ axis, and a coupling term (Fig.~\ref{fig:2D_reference}-A).
The eigenvalue spectrum of the Fokker-Planck operator for this system (eq.~\ref{eq:FP}) exhibits four dominant eigenvalues
at $\kappa_0^{2D} = 0$, $\kappa_1^{2D} = -5.98$, $\kappa_2^{2D} = -10.2$, and $\kappa_3^{2D} = -15.06$ (fig.~\ref{fig:2D_reference}.B). 
The corresponding eigenfunctions $l_0(x,y)$ to $l_3(x,y)$ are shown in (fig.~\ref{fig:2D_reference}-D). 
The eigenvector $l_0(x,y)$ is equal to the stationary density. 
Eigenvectors $l_1(x,y)$ and $l_2(x,y)$ describe slow transitions along the $x$ and $y$ axis, respectively. 
Eigenvector $l_4(x,y)$ represents a dynamic process which mixes $x$ and $y$, and is due to the coupling term in $V(x,y)$.
We constructed this reference solution by evaluating the SqRA (eq.~\ref{eq:rate_matrix_01}) on a quadratic grid (eq.~\ref{eq:rate_adjacent_rectangular}) with $N=50\times 50 = 2500$ cells on the space $[-6, 6] \times [-6, 6]$.

Additionally, we constructed a MSM on the same grid. We generated a time-discretized trajectory of $1 \times 10^8$ time-steps, with a time-step $\Delta t = 0.001$, integrating eq.~\ref{eq:sde} according to the Euler-Maruyama scheme \cite{Leimkuhler2015}.
The MSM has been constructed by counting transitions $C_{ij}(\tau)$ from cell $\Omega_i$ to cell $\Omega_j$ within a lag time $\tau$ varied in a range of [5:500] time-steps \cite{Prinz2011}.
Detailed balance has been enforced by symmetrizing the resulting $50\times50$-count matrix: 
$\mathbf{C}_{\mathrm{sym}}(\tau) = \mathbf{C}(\tau) + \mathbf{C}^{\top}(\tau)$, where $\mathbf{C}^{\top}(\tau)$ denotes the transpose of $\mathbf{C}(\tau)$.
The MSM transition matrix $\mathbf{T}(\tau)$ was obtained by row-normalizing $\mathbf{C}_{\mathrm{sym}}(\tau)$. 
The eigenvectors of the MSM transition matrix are defined as
$\mathbf{l}_i^{\top}\mathbf{T}(\tau) = \lambda_i(\tau) \mathbf{l}_i^{\top}$.
The MSM yielded the same dominant eigenvectors as the SqRA of the rate matrix. 
The MSM eigenvalues $\lambda_i(\tau)$  and the eigenvalues of the rate matrix can be interconverted by
\begin{eqnarray}
    \lambda_i(\tau) = \exp\left(\kappa_i\tau\right) &\Leftrightarrow&  \kappa_i = \frac{\ln(\lambda_i(\tau))}{\tau} \, .
\end{eqnarray}
and are in excellent agreement (Fig.~\ref{fig:2D_reference}-C).
The fact that the ratio $\frac{\ln(\lambda_i(\tau))}{\tau}$ does not vary with $\tau$ indicates that the MSM on this grid has a negligible projection error.
Since the SqRA-model and the MSM do not deviate from each other, we can assume that also the SqRA-model has a negligible projection error.
We will therefore use the SqRA-model on a regular grid with $N=2500$ cells as a reference solution for further tests.
%

%
%
%
%
%

\begin{figure*}[!ht]
  \begin{center}
  \includegraphics[scale=1]{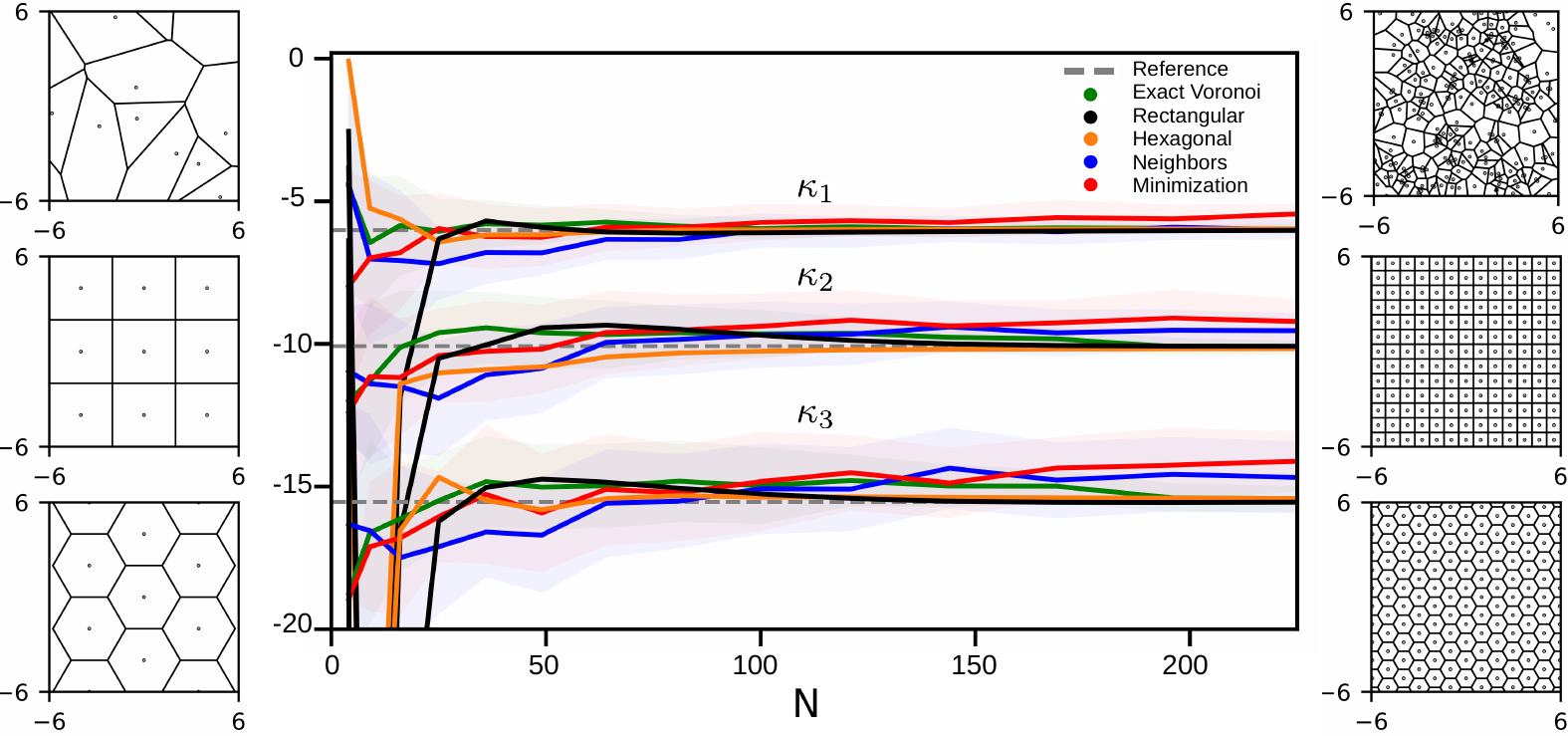}
    \caption{Accuracy  of  the  flux  estimated  by  discretizing  the  Laplacian: Reference (gray); Exact Voronoi (green);  Rectangular  grid (black); Hexagonal grid (orange); Neighbors (blue); Minimization (red). }
   \label{fig:2D_Laplacian_test}
  \end{center}
\end{figure*}

To assess whether the method to estimate the flux has influence in the accuracy of the SqRA of the rate matrix, we varied the number of grid cells from $N = 4$ to $N = 225$.
We constructed quadratic grids, hexagonal grids and arbitrary Voronoi grids.
For the arbitrary Voronoi grids, we randomly placed grid centers in the two-dimensional state space. 
To account for the variance in these randomly constructed grids, we constructed fifty different grids for each value of $N$ and constructed the corresponding rate matrix. 
In Fig.~\ref{fig:2D_Laplacian_test}, we report the mean and the variance of the dominant eigenvalues for Voronoi grids, 
that were calculated using the methods ``exact'' (green), ``neighbors'' (blue), and ``minimization'' (red).
Fig.~\ref{fig:2D_Laplacian_test} also shows the dominant eigenvalues for the quadratic grid calculated using the method ``rectangular'' (black) and the hexagonal grid calculated using the method ``hexagonal'' (orange), 
as well as the reference value for the eigenvalues (dashed).
The results for the Voronoi grids seem to converge faster than the results for the regular grid. 
The mean of the eigenvalues for the Voronoi grids is already reasonably accurate for $N=9$ or $N=16$ grid cells. 
Note however that the variance is sizeable at these low numbers of grid cells and that, depending on the exact location of the grid cells, the Voronoi results can also deviate considerably from the reference value.
For $N=25$ all five methods yield results that are close to the reference value, and the accuracy of all five methods increases only slowly with increasing $N$.
In fact, between $N=100$ and $N=225$ we do not find a significant improvement for any of the five methods.
This means that, at least for this potential energy function, 25 to 100 grid cells are sufficient to discretize 
the two-dimensional state space. 
This is an order of magnitude lower than previously reported discretizations of two-dimensionsal molecular states spaces that relied on the SqRA \cite{Rosta2014, Wan2016}.
For $N>100$ the eigenvalues obtained from regular grids are almost exactly equal to the reference values, 
whereas the results from Voronoi grids tend to overestimate the eigenvalues. 
This indicates that, if one is interested in a highly accurate estimate of the dominant eigenvalues, 
one should opt for a regular grid.


\newpage
\section{Conclusion}
We have derived, from the equation of the overdamped Langevin dynamics with constant potential, the expression of the flux $\Phi$ that appears in the SqRA formula \cite{Lie2013, Donati2018b}.
An analogous formula, was previously derived for the one-dimensional Smoluchowski equation discretized on a regular grid \cite{Bicout1998} and later used to estimate the diffusion coefficients of molecular systems projected on one-dimensional relevant coordinates \cite{Hummer2005, Schulz2017}.
Our result generalizes to the case of $N_D$-dimensional diffusive systems discretized on multidimensional arbitrary grids.
Moreover, we proposed and tested several methods which can be used to calculate the exact or the approximate value of the multiplicative factor $\Phi\frac{\mathcal{S}_{ij}}{\mathcal{V}_i}$ 
for different grid types.
We now have an approach in place that in principle allows us  
to calculate MSMs of molecular systems without running MD simulations.
The accuracy with which the dominant eigenvalues of the rate matrix can be estimated is similar for all methods.
But our analysis has shown that, depending on the grid and the method to estimate the flux, the relative and absolute computational costs of these three methods vary drastically. 
The entire computation of the SqRA of the Fokker-Planck equation, from discretization of the state space to the analysis of the dominant eigenvectors, consists of three steps: 
($i$) generate the adjacency matrix,
($ii$) calculate the rates $Q_{ij}$,
($iii$) calculate the eigenvalue and eigenvectors of the rate matrix.
On a regular grids, the generation of the adjacency matrix using the algorithm in the supplementary material is computationally cheap.
The factor $\Phi\frac{\mathcal{S}_{ij}}{\mathcal{V}_i}$ is essentially a constant, and the computational cost is dominated by the calculation of the eigenvectors.
Thus, calculating the SqRA on a regular grid is by far the most efficient approach, if one aims at discretizing the entire state space.
However, molecules at room temperature only access a small fraction of their state space, 
and the experience with MSMs has shown that Voronoi grids are useful for discretizing the accessible state space \cite{Prinz2011}.
We therefore do not yet want to rule out Voronoi grids.
We have compared three methods to calculate $\Phi\frac{\mathcal{S}_{ij}}{\mathcal{V}_i}$: 
an ``exact'' method that aims at calculating $\frac{\mathcal{S}_{ij}}{\mathcal{V}_i}$ numerically, and two approximate methods. 
The ``neighbor'' method is based on an already known interpolation scheme between all neighbors of a given cell $\Omega_i$. 
The ``minimization'' method is an approach that we proposed in this contribution, and is based on a comparison to the analytically known transition probability at constant potential.
On Voronoi grids, the construction of the adjacency matrix is computationally much more demanding than on regular grids, and for the approximate methods, the computational cost is dominated by the construction of the adjacency matrix. 
However, with the ``exact'' method the calculation of $\frac{\mathcal{S}_{ij}}{\mathcal{V}_i}$ is the most costly step,
increasing the computational cost of the entire calculation by an order of magnitude. 
Taking into account that this method only slightly improves the accuracy of the MSM, the ``exact'' method is not suited for an actual application. 
With a few seconds computing time on a single compute server, we could reach $\mathcal{O}(10^6)$ grid cells for regular grids, while we need a fews days to reach $\mathcal{O}(10^3)$ for Voronoi grids. 
Using the ``neighbors'' or the ``minimization'' method, $\mathcal{O}(10^4)$ are within reach for Voronoi grids, and moving the calculation to high-performance compute clusters or GPUs will likely push the limit to $\mathcal{O}(10^5)$ states.
With grids of this size, the SqRA becomes useful for small molecules.
However, a brute-force discretization of the entire state space of larger molecules would require even larger grids. 
There are two possible remedies: 
($i$) one discretizes only the accessible state space, or
($ii$) one projects the dynamics on a lower-dimensional space and discretizes this space.
In the first approach, the accessible state space will best be formulated in terms of internal coordinates. 
Then the distances between grid cells and potentially also the cell volumes have to be transformed accordingly. 
In the second approach, one needs to calculate the free-energy surface and the position-dependent flux on the low-dimensional space,
for which MD simulations are needed \cite{Risken1989, Hanggi1990, Bicout1998, Hummer2005}.
Additionally, one needs to adjust the SqRA to account for the position-dependent flux. 
For one-dimensional regular grids, the adjusted SqRA rates are reported in refs.~\onlinecite{Bicout1998} and \onlinecite{Hummer2005}.
Note that the second approach is currently not limited by the computational cost for the SqRA, 
but by the computational cost for the MD simulations.
In future work we compare these two approaches, and apply the SqRA to molecular systems.

\section{Supplementary material}
See supplementary material for the example script to construct an adjacency matrix for $N_D$-dimensional systems on hyper-cubic grids and to construct the rate matrix $\mathbf{Q}$ using the  "rectangular" method. 

\begin{acknowledgments}
This research has been funded by the Deutsche Forschungsgemeinschaft (DFG, German Research Foundation) under Germany´s Excellence Strategy – EXC 2008/1 (UniSysCat) – 390540038, and through grant CRC 1114 ``Scaling Cascades in Complex Systems'', project B05 ``Origin of the scaling cascades in protein dynamics''. B.G.K.~is grateful for a writing retreat funded by ``Die Junge Akademie''.
\end{acknowledgments}

\appendix
\newpage
\section{Estimating the flux from Fick's laws}
In the following, we provide an alternative derivation of the quantity $\Phi$ that appears in eq.~\ref{eq:rate_matrix_01}.
The Fokker-Planck equation (eq.~\ref{eq:FP}) can be written in the form of a continuity equation \cite{Risken1989}, 
which at constant potential energy function reduces to Fick's second law of diffusion \cite{Engels2014}
\begin{eqnarray}
\partial_t \rho(x,t)  
&=& \frac{\sigma^2}{2}  \Delta \rho(x,t) = -\nabla \mathbf{J} \, ,
\label{eq:FPEflux}
\end{eqnarray}
where the flux is given by 
\begin{eqnarray}
    \mathbf{J}(x) &=& -\frac{\sigma^2}{2} \nabla \rho(x,t) \, .
\end{eqnarray}
To discretize the $\nabla \mathbf{J}$ on a Voronoi grid, we apply the Gauss theorem, and discretize the surface integral along the sides of the
Voronoi cell $\Omega_j$
\begin{eqnarray}
 \left. \nabla \mathbf{J} \right\vert_{x=x_j} 
 & = & \lim_{\mathcal{V}_j\rightarrow 0} \frac{1}{\mathcal{V}_j} \oint_{\partial\Omega_j \partial \Omega_i}  \mathbf{J}(z) \, \mathrm{d}\mathcal{S}(z)
 = \lim_{\mathcal{V}_j\rightarrow 0} \frac{1}{\mathcal{V}_j} \sum_{i\sim j} \mathbf{J}_{ji}  \, \mathcal{S}_{ji}\, ,
 \end{eqnarray}
where
\begin{eqnarray}
    \mathbf{J}_{ji} = -\frac{\sigma^2}{2}  \left.\nabla \rho(x,t) \right\vert_{x = x_j} \cdot \mathbf{n}_{ji}
\end{eqnarray}
is the flux in direction $\mathbf{n}_{ji} = \frac{x_i - x_j}{h_{ji}}$.
Using the same finite difference as in eq.\ref{eq:rho_finite_difference}, we obtain 
\begin{eqnarray}
\partial_t  \rho(x_j, t)
&=& - \nabla \mathbf{J} 
= \lim_{\mathcal{V}_j\rightarrow 0} \frac{1}{\mathcal{V}_j} \sum_{i\sim j}  \frac{\sigma^2}{2}  \frac{\rho(x_i, t) - \rho(x_j, t)}{h_{ji}} \cdot \mathcal{S}_{ji} \, .
\label{eq:discFPE2}
\end{eqnarray}
To convert the continuous probability density evaluated at the cell centers $\rho(x_j,t)$ to discrete probability defined on finite cell volumes
we use the relation $\rho_j(t) = \int_{\Omega_i} \rho(x,t) \, \mathrm{d}x \approx \rho(x_j,t) \mathcal{V}_j$.
We obtain
\begin{eqnarray}
\partial_t \rho_j(t)
&=& \frac{\sigma^2}{2} \sum_{i\sim j} \left( \frac{\rho_i(t)}{\mathcal{V}_i} - \frac{\rho_j(t)}{\mathcal{V}_j} \right) \frac{1}{h_{ji}} \cdot \mathcal{S}_{ji} 
\end{eqnarray}
which is identical to eq.~\ref{eq:discFPE}.
And thus also from the view-point of the continuity equation and Fick's laws of diffusion the flux is given as 
\begin{equation}
    \Phi = \frac{\sigma^2}{2 h_{ij}} \,.
\label{eq:flux02}    
\end{equation}{}
%
%
\section{The limit of infinitesimally small cells}
\label{app:convergenceToFP}
We show that the SqRA on a (hyper-)cubic grid  converges to the Fokker-Planck equation in the limit of infinitesimally small cells.
On these grids the cell length $h$ is the same in each grid dimension, but the extension to rectangular grids is straight forward.

The rates between adjacent cells are given by eq.~\ref{eq:rate_adjacent_rectangular}
\begin{equation}
\label{eq:Q_ij}
Q_{ij}  = \frac{\sigma^2}{2} \,\frac{1}{h^2}\sqrt{\frac{\pi(x_j)}{\pi(x_i)}}
        = \frac{\sigma^2}{2\,h^2} \exp \left(-\beta \frac{V_j - V_i}{2} \right)\, ,
\end{equation}
which yields the following the master equation (eq.~\ref{eq:FP_discretized02})
\begin{eqnarray}
    \partial_t \rho_j(t) 
    &=& \frac{\sigma^2}{2\,h^2} \sum_{i \sim j}  \left[\rho_i(t) \exp \left(-\beta \frac{V_j - V_i}{2} \right) - \rho_j(t)  \exp \left(-\beta \frac{V_i - V_j}{2} \right)\right] \, .
\end{eqnarray}
We replace the exponential function in eq.~\ref{eq:Q_ij} by its first-order Taylor expansion,
$\exp \left(-\beta \frac{V_j - V_i}{2} \right) = 1 - \beta \frac{V_j - V_i}{2} + \mathcal{O}\left((V_j - V_i)^2\right)$,
and obtain
\begin{eqnarray}
    \partial_t \rho_j 
    &\approx&  \frac{\sigma^2}{2\,h^2} \sum_{i \sim j }  
    \left[\rho_i \left( 1 - \beta \frac{V_j - V_i}{2}\right) 
        - \rho_j \left( 1 - \beta \frac{V_i - V_j}{2} \right)\right] 
        \cr  
   &\approx&  \frac{\sigma^2}{2\,h^2} \sum_{i \sim j}  
    \left[(\rho_i - \rho_j) - \rho_i\beta \frac{V_j - V_i}{2} + \rho_j \beta \frac{V_i - V_j}{2} \right] \, ,
\end{eqnarray}
where we omitted the $t$-dependence of $\rho_i(t)$ and $\rho_j(t)$ for the sake of brevity.
Next we write $V_j - V_i = - (V_i - V_j)$ 
and substitute $\rho_j \beta \frac{V_i - V_j}{2} = \rho_j \beta V_i - V_j -\rho_j \beta \frac{V_i - V_j}{2}$
\begin{eqnarray}
    \partial_t \rho_j 
   &\approx&  \frac{\sigma^2}{2\,h^2} \sum_{i \sim j }
    \left[(\rho_i - \rho_j) + \rho_i\beta \frac{V_i - V_j}{2} + \rho_j \beta (V_i - V_j)- \rho_j \beta \frac{V_i - V_j}{2} \right] \cr
   &\approx&  \frac{\sigma^2}{2\,h^2} \sum_{i \sim j }
    \left[(\rho_i - \rho_j) +  \rho_j \beta (V_i - V_j) + (\rho_i- \rho_j) \beta \frac{V_i - V_j}{2} \right] \, .
\label{eq:FP_discretized03}       
\end{eqnarray}
We revover the continous probability density $\rho(x,t)$ from the 
discrete probabilities using the relation 
$\rho_i(t) = \rho(x_i,t)\, \mathcal{V}_i = \rho(x_i,t)\, h^n$, 
and the relation for the potential energy function 
$V_i = V(x_i)$.
The cell volume $h^n$ appears linearly on both sides of the equation, and cancels: 
\begin{eqnarray}
    \partial_t \rho(x_j) 
   &\approx&  \frac{\sigma^2}{2\,h^2} \sum_{i \sim j } 
            \left[(\rho(x_i) - \rho(x_j)) +  \rho(x_j) \beta (V(x_i) - V(x_j)) \right.\cr
    &&      \left.  + (\rho(x_i)- \rho(x_j)) \beta \frac{V(x_i) - V(x_j)}{2} \right] \, ,
\label{eq:FP_discretized04}     
\end{eqnarray}
where $x_i$ and $x_j$ are the (still discrete) cell enters, 
and we omit the $t$-dependence of $\rho(x,t)$ for the sake of brevity.
We remind the reader that $\sum_{i \sim j }$ denotes a sum over all cells $\Omega_i$ which are adjacent to cell $\Omega_j$.
On a regular grid, every cell $\Omega_j$ has two neighbors in each grid dimension, 
which are centered at $x_j + h\cdot n_k$ and $x_j - h\cdot n_k$, 
where $n_k$ is the unit vector pointing in direction $k$, and
$h\cdot n_k$ is the lattice vector along the $k$th dimension.
We will now sort the sum over adjacent cells according to grid dimension $k$ and will take the limit $h \rightarrow 0$ to recover the differential equation.
With this approach, the first term in eq.~\ref{eq:FP_discretized04} becomes
\begin{eqnarray}
    \lim_{h\rightarrow 0} \frac{1}{h^2} \sum_{i \sim j }(\rho_i - \rho_j) 
    &=&  \lim_{h\rightarrow 0} \frac{1}{h^2} \sum_{k=1}^n\left[\rho(x_j - h_k) - \rho(x_j) + \rho(x_j + h_k) - \rho(x_j)  \right] \cr
    &=&  \lim_{h\rightarrow 0}
        \sum_{k=1}^n
        \frac{\rho(x_i-h_k) - 2\rho(x_i) + \rho(x_i + h_k)}{h^2} \cr
    &=& \sum_{k=1}^n 
         \partial_k^2 \rho(x_j) \cr
    &=& \Delta \rho(x_j,t)   \, ,  
\end{eqnarray}
where $\partial_k$ denotes the derivative with respect to the $k$th 
dimension. 
Similarly, 
\begin{eqnarray}
    \lim_{h\rightarrow 0} \frac{1}{h^2}\sum_{i \sim j }^n  
    \rho(x_j) \beta (V(x_i) - V(x_j))
    &=& \beta \rho(x_j,t) \Delta V(x_j) \, .
\end{eqnarray}
The third term in eq.~\ref{eq:FP_discretized04} has the following limit
\begin{eqnarray}
    &&  \lim_{h\rightarrow 0} \frac{1}{h^2} \sum_{k=1}^n
        (\rho(x_i)- \rho(x_j)) \beta \frac{V(x_i) - V(x_j)}{2}\cr
    &=& \lim_{h\rightarrow 0} \frac{1}{h^2} \sum_{k=1}^n
    \left[
    \frac{\beta}{2} \frac{\rho (x_j+h_k) - \rho(x_j)}{h_k} \frac{V(x_j+h_k) - V(x_i)}{h_k} +
    \frac{\beta}{2} \frac{\rho (x_j-h_k) - \rho(x_j)}{h_k} \frac{V(x_j-h_k) - V(x_i)}{h_k}
    \right] \cr
    &=&\beta \sum_{k=1}^n \partial_k \rho(x_j) \, \partial_k V(x_j)  \cr
    &=& \beta \nabla \rho(x_j,t) \nabla V(x_j)\, .
\end{eqnarray}
In the limit $h \rightarrow 0$ eq.~\ref{eq:FP_discretized03} becomes
\begin{equation}
\label{eq:me4}
\partial_t \rho(x,t)  = \frac{\sigma^2}{2} \left[ \Delta \rho(x,t) + \beta  \rho(x,t) \Delta V(x) + \beta \nabla \rho(x,t)  \nabla V(x) \right] \, .
\end{equation}
Applying the product rule and using $\beta = \frac{2}{\sigma^2}$, 
we obtain the Fokker-Planck equation as stated in eq.~\ref{eq:FP}:
\begin{equation}
\partial_t \rho(x,t)  = \frac{\sigma^2}{2}  \Delta \rho(x,t) + \nabla \left(\rho(x,t) \cdot \nabla V(x) \right)   \, .
\end{equation}
%
%
%
%
%
%
\section{Estimating $\Phi_{\mathrm{grid}}$ via Taylor expansion of the propagator.}
\label{app:FluxViaTaylor}
The starting point for the derivation is eq.~\ref{eq:transitionMatrix}. 
We then express the exponential function in terms of its Taylor series 
and truncate the series after the linear term
\begin{eqnarray}
\mathbf{T}(\tau)&=& \exp(\tau\mathbf{Q})  = \exp(-\tau\, \Phi_{\mathrm{grid}}\mathbf{L}) 
= \sum \frac{(-\tau\, \Phi_{\mathrm{grid}}\mathbf{L})^n}{n!} \approx \mathbf{1} - \tau\, \Phi_{\mathrm{grid}}\mathbf{L}
\end{eqnarray}
This approximation is valid at small values of $\tau$, and yields the following approximate expression for the rate matrix 
\begin{eqnarray}
	\mathbf{Q} = -\Phi_{\mathrm{grid}}\mathbf{L} 
	&=& \frac{1}{\tau} \left(\mathbf{T}(\tau) - \mathbf{1}  \right) \, ,
\end{eqnarray}
where $\mathbf{1}$ is the identity matrix.
For adjacent cells $L_{ij}=-1$, and $\left[\mathbf{1}\right]_{ij}=0$, and we obtain the following equation for $\Phi_{\mathrm{grid}}$
\begin{eqnarray}
	\Phi_{\mathrm{grid}} \approx  \frac{1}{\tau} T_{ij}(\tau) 
	&=& \frac{1}{\tau} \left(\sqrt{\frac{1}{2\pi\sigma^2 \tau}}\right)^n \exp\left(-\frac{h_{ij}^2}{2\sigma^2 \tau} \right)  \mathcal{V} \cr
	&=& \Phi_{\mathrm{grid},\, \mathrm{approx}}(\tau)\, ,
\label{eq:fluxViaTransProb02}	
\end{eqnarray}
where we used eq.~\ref{eq:transProb02} to express $T_{ij}(\tau)$.
Given the parameters $n$, $\sigma$, $h_{ij}$, $\mathcal{V}$, and setting $\tau$ to some fixed value, one can in principle calculate an approximation of $\Phi_{\mathrm{grid}}$.
However, $\Phi_{\mathrm{grid},\, \mathrm{approx}}(\tau)$ is very sensitive to $\tau$. It is positive everywhere. 
For $\tau \rightarrow 0$ and for $\tau \rightarrow \infty$, it approaches zero, and in between it has a maximum which, depending on the other parameters, can be very steep. 
Thus, choosing $\tau$ arbitrarily leads to very inconsistent results. 
Let us instead choose the value of $\tau$ at which $\Phi_{\mathrm{grid},\, \mathrm{approx}}(\tau)$ reaches its maximum as the optimal value for $\tau$.
The derivative of $\Phi_{\mathrm{grid},\, \mathrm{approx}}(\tau)$ with respect to $\tau$ is
\begin{eqnarray}
    \frac{d}{d\tau} \Phi_{\mathrm{grid},\, \mathrm{approx}}(\tau)
    &=& - \frac{1}{\tau} \left[1  + \frac{n}{2}  -  \left(\frac{h_{ij}^2}{2\sigma^2 \tau} \right)\right] 
    \cdot \Phi_{\mathrm{grid},\, \mathrm{approx}}(\tau) \, .
\end{eqnarray}
Setting $\frac{d}{d\tau} \Phi_{\mathrm{grid},\, \mathrm{approx}}(\tau) = 0$ and solving for $\tau$ yields
\begin{eqnarray}
    \tau_{\rm opt} &=& \left(1 + \frac{n}{2}\right)^{-1}\frac{h_{ij}^2}{2\sigma^2} \, .
\label{eq:tauOpt}    
\end{eqnarray}
Inserting eq.~\ref{eq:tauOpt} into eq.~\ref{eq:fluxViaTransProb02} yields
\begin{eqnarray}
    \Phi_{\mathrm{grid},\, \mathrm{approx}}(\tau_{\rm opt})
    &=& \frac{\left(n+2\right)\sigma^2}{h_{ij}^2} \cdot \frac{1}{h_{ij}^n} \cdot
        \left(\frac{\left(1 + \frac{n}{2}\right)}{\pi}\right)^{n/2} \cdot \exp\left(-\left(1 + \frac{n}{2}\right) \right) \cdot \mathcal{V} \, .
\label{eq:phiTaylor}        
\end{eqnarray}
To test whether eq.~\ref{eq:phiTaylor} is a useful approximation, we compare $\Phi_{\mathrm{grid},\, \mathrm{approx}}(\tau_{\rm opt})$ 
to the grid flux on a hyper-cubic grid which is given by eq.~\ref{eq:rate_adjacent_rectangular} as $\Phi_{\mathrm{grid},\, \mathrm{hyper-cube}}= \frac{1}{2}\frac{\sigma^2}{h_{ij}^2}$.
On a  hyper-cubic grid $\mathcal{V} = h_{ij}^n$, and eq.~\ref{eq:phiTaylor} simplifies to
\begin{eqnarray}
    \Phi_{\mathrm{grid},\, \mathrm{approx}}(\tau_{\rm opt})
    &=& \left(n+2\right) \cdot
        \left(\frac{\left(1 + \frac{n}{2}\right)}{\pi}\right)^{n/2} \cdot \exp\left(-\left(1 + \frac{n}{2}\right) \right) \cdot \frac{\sigma^2}{h_{ij}^2} \, .
\label{eq:phiTaylorHyperCube}        
\end{eqnarray}
Note that eq.~\ref{eq:phiTaylorHyperCube} scales correctly  with $\sigma$ and $h_{ij}$. 
%
For $n=1, 2, 3$ and $4$, the ratio $\Phi_{\mathrm{grid},\, \mathrm{approx}}(\tau_{\rm opt}) / \Phi_{\mathrm{grid},\, \mathrm{hyper-cube}}$  is respectively equal to $ 0.92, 0.68, 0.58$ and $0.54$; for $n>4$ the ratio grows exponentially as the term $\left(\frac{\left(1 + \frac{n}{2}\right)}{\pi}\right)^{n/2}$ in eq.~\ref{eq:phiTaylorHyperCube} dominates all other terms.
Thus, $\Phi_{\mathrm{grid},\, \mathrm{approx}}(\tau_{\rm opt})$ cannot be used as a valid approximation of the characteristic flux of the grid.
Since eq.~\ref{eq:phiTaylor} likely shows a similar behaviour for arbitrary Voronoi grids, we do not recommend using it, and have not included it in our analysis in the main part of the publication.
%


\bibliographystyle{unsrt}
\bibliography{references.bib}


\end{document}